# Feature extraction with mel scale separation method on noise audio recordings

Roy Rudolf Huizen, Florentina Tatrin Kurniati
Faculty of Informatics and Computer, Institut Teknologi dan Bisnis STIKOM Bali, Bali, Indonesia

| Article Info | ABSTRACT |
|---|---|
|  | This paper focuses on improving the accuracy of noise audio recordings. High-quality audio recording, extraction using the mel frequency cepstral coefficients (MFCC) method produces high accuracy. While the low-quality is because of noise, the accuracy is low. Improved accuracy by investigating the effect of bandwidth on the mel scale. The proposed improvement uses the mel scale separation methods into two frequency channels (MFCC dual-channel). For the comparison method using the mel scale bandwidth without separation (MFCC single-channel). Feature analysis using k-mean clustering. The data uses a noise variance of up to -16 dB. Testing on the MFCC single-channel method for -16 dB noise has an accuracy of 47.5%, while the MFCC dual-channel method has an accuracy better of 76.25%. The next test used adaptive noise-canceling (ANC) to reduce noise before extraction. The result is that the MFCC single-channel method has an accuracy of 82.5% and the MFCC dual-channel method has an accuracy better of 83.75%. High-quality audio recording testing for the MFCC single-channel method has an accuracy of 92.5% and the MFCC dual-channel method has an accuracy better of 97.5%. The test results show the effect of mel scale bandwidth to increase accuracy. The MFCC dual-channel method has higher accuracy.<br> |

*Corresponding Author:*

Roy Rudolf Huizen
Faculty of Informatics and Computer
Institut Teknologi dan Bisnis STKOM Bali
Jl. Raya Puputan No.86, Denpasar, Bali 80234, Indonesia
Email: roy@stikom-bali.ac.id

## 1. INTRODUCTION

Users of digital devices are increasing. This increase is influenced by relatively low prices, free applications, and fast internet access. This brings changes and habits. On previous communication devices using the telephone, switch to applications (WhatsApp and Telegram) [1], [2]. This change attracted the attention of law enforcement by starting to use digital evidence to prove a case [3], [4]. Digital evidence is a breakthrough to find out someone's involvement or activity, an example of digital evidence is recorded conversations. Conversation can be defined as interactive communication between individuals. Recording a conversation will clarify the chronology of an event [5].

The audio records are useful for knowing an event. The process is by converting the audio frequency using a microphone into an electrical signal and saving it in the form of a file [6]. Audio tapping by the authorities is an activity to record without being noticed. Wiretapping recordings are not necessarily good quality, because they can't choose the environmental situation when the tapping takes place [6], [7]. Noisy environment, wiretapping recording is noise, the audio recording quality is low [7]. This affects the accuracy of the information contained in the recording [5], [8]. Information on a audio record includes individual profiles, locations, and chronology of an event [9]. Obtaining information by extracting word samples, to obtain features for identification [10], [11]. The process is by matching the features of evidence and comparison. Similarity of features means that the audio records come from the same individual [12]. The similarity of features is calculated from the high value of accuracy. The extraction method is a factor that affects the accuracy value. The mel frequency cepstral coefficients (MFCC) method is a reliable method with high accuracy for high-quality audio recordings [13]. The accuracy rate is more than 90% [14], [15]. The high accuracy of the MFCC method is due to the mel scale which has characteristics similar to human hearing [11]. The word sample if extracted by the MFCC method, the value of the frequency component will match the characteristics of the mel scale. The resulting features represent human hearing. Mel scale sensitivity is one of the factors that increase the accuracy value. This method is superior for the identification of high-quality audio recording.

Identification of audio recordings has time intervals of up to tens of years with comparison audio recordings (aging-factor), accuracy decreases due to changes in frequency components [16]. The same applies to audio recordings with noise. Comparing noise word samples causes a decrease in accuracy due to changes in the frequency component. Each word sample consists of several frequency components, namely the fundamental frequency and the resonant frequency. The wiretapping audio recording contains noise. The frequency components are the audio frequency and the noise frequency. So that the features obtained consist of a combination of the word sample frequency with noise. If tested with a comparison sample, the accuracy is low [12]. These results lead to inaccurate analysis [17]. In order to increase the accuracy value of low-quality wiretapping recordings due to noise, by developing an extraction method.

## 2. RESEARCH METHOD

In this study, the proposed improvement is the mel scale separation method, before extracting the word sample by dividing it into two frequency parts (MFCC dual-channel). The separation is based on linear human hearing characteristics for frequencies below 1 kHz and logarithmic for frequencies above 1 kHz. The stages of the MFCC dual-channel method as shown in Figure 1. The research data uses a sample of words with noise, obtained from conversation recordings by adding random noise using a computer. A random noise variant is used from low to high.

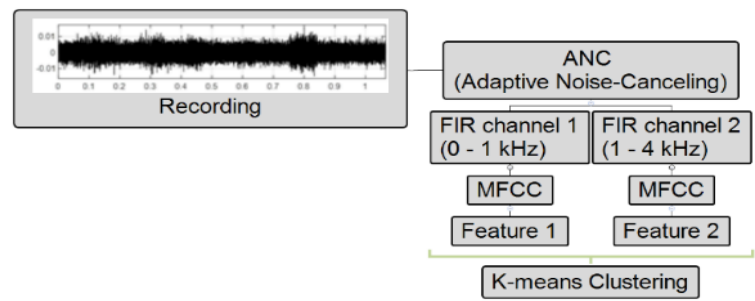

Figure 1. Identification stage with MFCC dual-channel

### 2.1. Adaptive noise-canceling

The noise reduction method in this study uses adaptive noise-canceling (ANC) with the least mean square (LMS) algorithm [18], [19]. This method has a simple and reliable structure [20], [21]. The structure of the LMS algorithm is shown in Figure 2(a). Identification of noise audio recordings, the first step is to reduce noise before extraction. This process requires reference noise ($X_k$). Input ($X_k$) as reference noise is processed by an adaptive linear combiner with a single input, as shown in Figure 2(b) [19], [22].

The reference noise as in (1) becomes the input of the adaptive linear combiner. The reference noise $X_{lk-1}$ passes the delay time ($z^{-1}$), the value of $X_{lk-1}$ is affected by the adaptively changing weight ($w_k$). The output of the adaptive linear combiner is shown in (2).

$$X_{lk-1} = [x_k \quad x_{k-1} \quad x_{k-2} \ldots x_{k-L}]^T \tag{1}$$

$$y_k = \sum_{l=0}^{L} w_{lk} x_{lk-1} \tag{2}$$

The adaptive linear combiner output ($y_k$) will be corrected through iterations that continue until the mean squared error (MSE) is minimal, as shown in (3).

$$\xi = E\,[e_k^2] \tag{3}$$

Minimum MSE means that the noise in the word sample is reduced [23]. The noise reduction is determined by the accuracy of the weight value on the adaptive linear combiner. Each stage of iteration, the amount of error is corrected to the desired limit. The reduction process error in ANC is shown in (4).

$$ek = dk - XkWTk \tag{4}$$

Based on (4), the determining factor for the iteration speed and the minimum MSE is determined from the accuracy of the weight value. In order to obtain the maximum reduction value and minimum iteration, the Steepest Descent method is used, by entering the step size ($\mu$), as shown in (5).

$$W_{k+1} = W_k - \mu \frac{d\xi}{dW} \tag{5}$$

In (5) the correction of the weight value $\frac{d\xi}{dW}$ using a zero gradient is shown in (6). Changes in the weight value (subtracting or adding) are affected by the reference noise value and the magnitude of the ANC error, shown in (6). Based on (5) and (6), the weight for noise reduction obtained by (7) is known as the LMS algorithm [20], [24].

$$\frac{d\xi}{dW} = -2\varepsilon_k X_k \tag{6}$$

$$W_{k+1} = W_k + 2\mu\varepsilon_k X_k \tag{7}$$

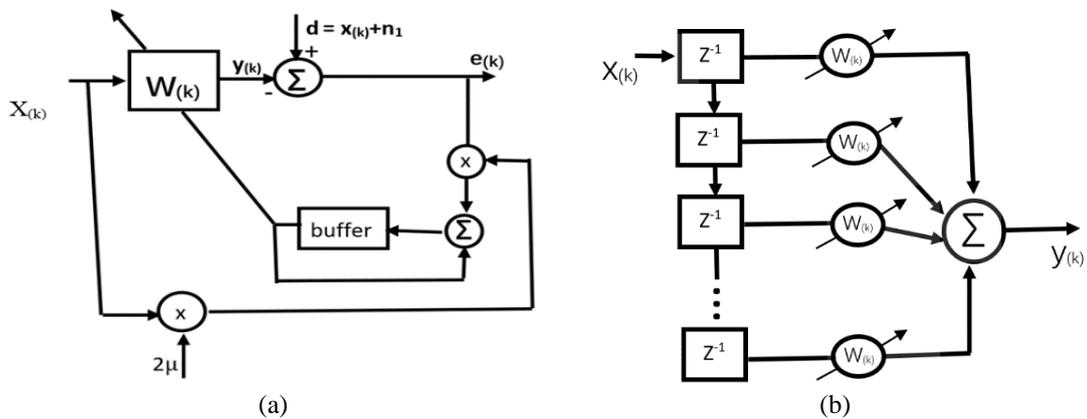

Figure 2. Adaptive noise-canceling with LMS algorithm; (a) LMS algorithm and (b) adaptive linear combiner with a single input

### 2.2. MFCC dual-channel

Extraction in the MFCC dual-channel method begins by separating the word sample into two parts. On channel 1, filtering uses low pass filter (LPF) for frequencies less than 1 kHz. While channel 2 uses band pass filter (BPF) with a frequency between 1 kHz to 4 kHz. Impulse responses for low pass filter (LPF), high pass filter, and band pass filter (BPF) refer to (8), (9), and (10).

$$h[n] = \frac{\Omega_0}{\pi}\,sinc\,(\Omega_0 n) \tag{8}$$

$$h[n] = sinc(\pi n) - \frac{\Omega_0}{\pi}\,sinc\,(\Omega_0 n) \tag{9}$$

$$h_{bp} = h_{LPH} - h_{LPL} \tag{10}$$

The determination of the bandwidth of both channels refers to the linear and logarithmic characteristics of human hearing at a certain frequency [16]. The noise recording (sample word) is shown in Figure 3(a). Records are separated by finite impulse response (FIR) [25], [26]. The result shown in Figure 3(b) for channel 1 shows that the noise tends to decrease compared to channel 2 in Figure 3(c). While the recording of noise reduced by ANC is shown in Figure 3(d). The results of the separation are shown in Figure 3(e), for channel 1 and Figure 3(f) for channel 2.

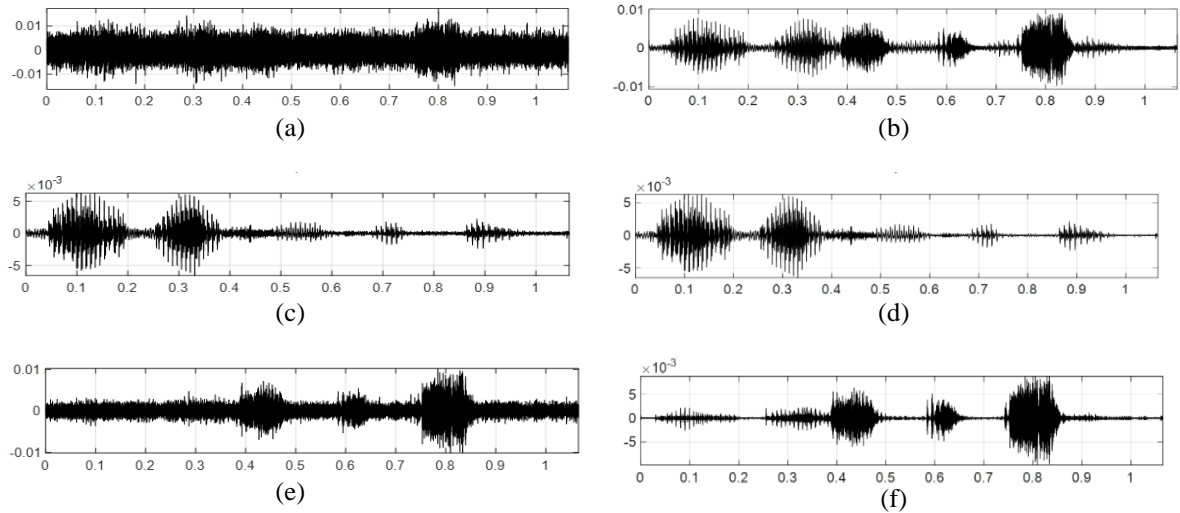

Figure 3. Words sample; (a) word sample with noise; (b) word sample with a noise, output of channel 1; (c) word sample with a noise, output of channel 2; (d) word sample with noise, reduced using ANC; (e) noise-reduced with ANC, output channel 1; (f) noise-reduced with ANC, output channel 2

The word samples that have been separation are then extracted using MFCC with the following steps; frame blocking, windowing, fast Fourier transform (FFT), mel scale filterbank, discrete cosine transform (DCT) and mel frequency cepstral coefficients [12]. The frequency component contained in the sample words in each channel, the frequency value is obtained by dividing the sample words into several frames. The frame length is set so that each frame has a frequency value. Setting the frame length using (11). If the frame length is N and shifted by M then each frame has M overlap. The frame blocking process is shown in Figure 4.

$$(1:n) = s'(N+M(l-1)) \qquad (11)$$

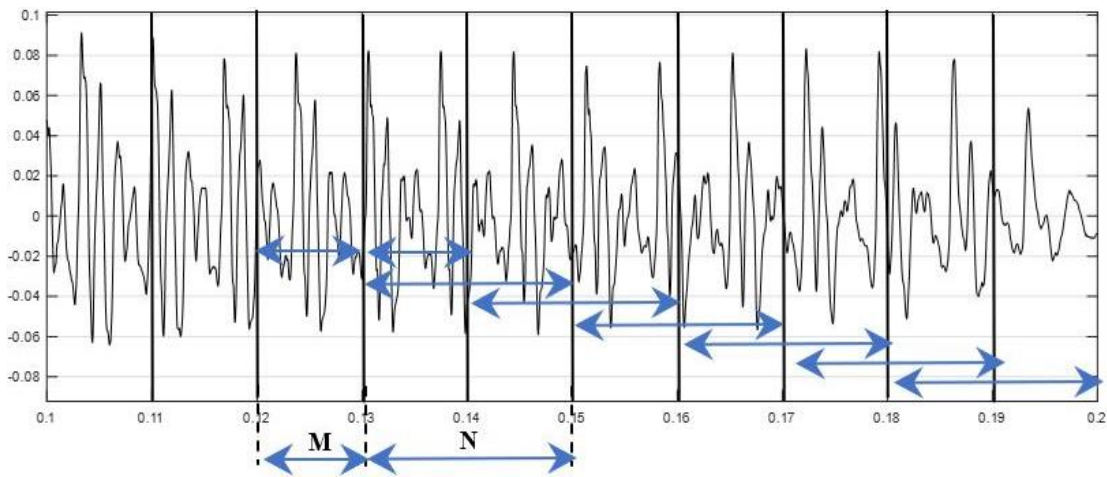

Figure 4. Frames blocking

Each frame is windowed, this process aims to reduce spectrum leakage due to the low sampling process. The type of windowing used is Hamming, shown in (12) and (13);

$$(n) = 0.54 - 0.46 \cos(2\pi n/(N-1)) \tag{12}$$

$$x(n) = xl * w(n) \tag{13}$$

each frame after windowing is calculated the frequency value, using discrete Fourier transform (DFT) (converting from time domain to frequency). A total of N DFT data is calculated using FFT, to determine the frequency value, (14) is used.

$$X_{(k)} = \sum_{n=1}^{N} x_{(n)} \cdot e^{-j\left[\frac{2\pi k n}{N}\right]} \tag{14}$$

$$fmel\_ch1,2 = 2595 * \log 10 \left[1 + \frac{f_{lin}}{700}\right] \tag{15}$$

On channel 1 the mel scale is for scaling frequencies below 1 KHz and channel 2 is for scaling frequencies between 1 to 4 KHz. The frequency of the process is the mel frequency ($f_{mel\_ch1,2}$) [27]. The characteristics of the mel scale on channels 1 and 2 follow the principle that not all follow a linear pattern, so that each channel follows a linear and exponential pattern. The frequency is then on the mel scale with (15). The width of the mel scale plane of channel 1 and channel 2 is shown in Figure 5.

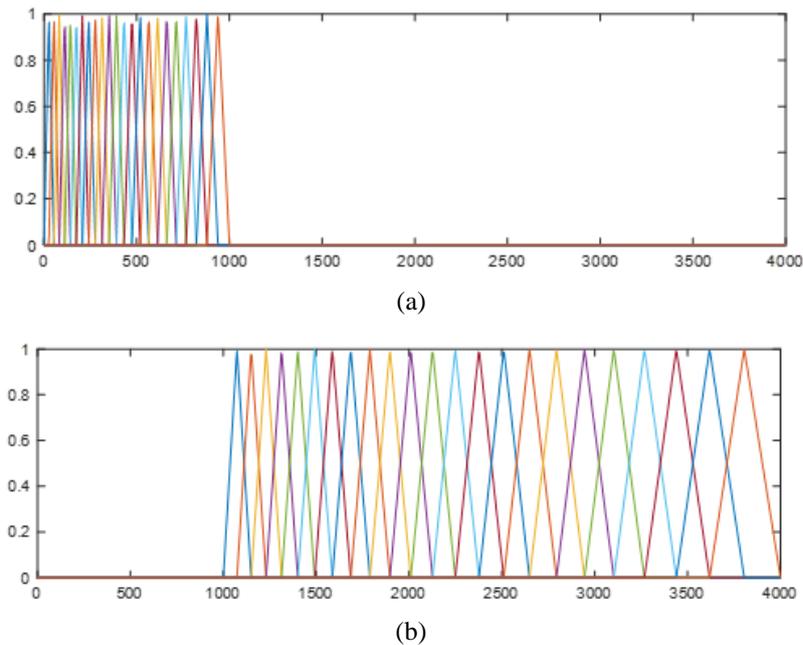

(a)

(b)

Figure 5. Mel scale with filterbank on MFCC dual-channel; (a) channel 1, and (b) channel 2

The next step, the mel frequency value is then calculated for the log mel frequency using (16). Furthermore, with (17) the mel cepstrum coefficient value is obtained which is the feature ceptral coefficient of the word sample for each channel.

$$a_{mfc_n^m\_ch12,(k)} = ln(\sum_{k=1}^{K} |X_{n\_}ch1,2(k)|^2 H_m K\_ch1,2) \tag{16}$$

$$C_n^q\_ch1,2 = \sum_{m=1}^{P} a\_mfc_n^m\_ch1,2 \cos\left[m\left(q - \frac{1}{2}\right)\frac{\pi}{P}\right] \tag{17}$$

From the extraction, the features obtained are shown in channels 1 and 2 shown in Figure 6.

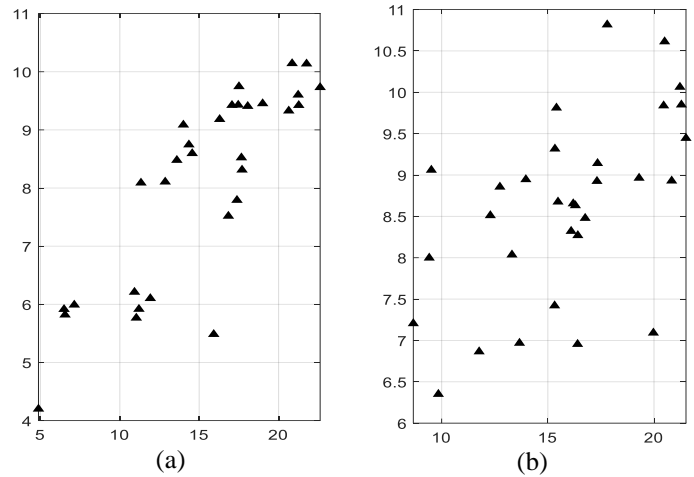

Figure 6. Features the result of extracting the noise word sample and reduce by ANC; (a) channel 1 features and (b) channel 2 features

### 2.3. K-mean clustering

The feature analysis on each channel uses k-mean clustering, the data for each feature in the word sample is calculated for the cluster center [23], [28]. Identification determines whether a sample of words is identical or not by comparing the distance to the center of the cluster. Identical if the centroid of the word sample has a high proximity value to the comparison word sample and not identical if the two samples are relatively far apart between the centroids. The steps for k-mean clustering are as follows;

− Step 1: Determine the starting point K randomly from the feature data set. The next iteration uses (18).

$$C_i = \frac{1}{M}\sum_{j=1}^{M} x_j \qquad (18)$$

− Step 2: Each feature data with initial centroid is calculated using (19).

$$D(x_i, C_i) = \sqrt{\sum_{j=1}^{q}(x_{ij} - C_{ij})^2} \qquad (19)$$

− Step 3: Update the centroid based on the calculation of stage 2.
− Step 4: The second step is repeated until there is no change in the centroid value or it is stable.

The results of identification with the extraction method are calculated for the accuracy value using (20).

$$\text{Accuracy} = \frac{Tp+TN}{Tp+TN+FP+FN} \qquad (20)$$

To determine the performance of the MFCC dual-channel and MFCC single-channel methods by calculating the values of true positive (TP), true negative (TN), false positive (FP), false negative (FN) [7]. TP is a sample of words stating true, and the test results are identical. TN is a sample of words stating true, and the test results are non-identical. For FP, the sample word is false, and the test states identical. Meanwhile, FN is a sample of false words, and the test results state that they are non-identical. To determine the performance of the MFCC dual-channel and MFCC single-channel methods by calculating the accuracy value by comparing TP and TN with TP, TN, FP, FN as shown in (20).

## 3. RESULTS AND DISCUSSION

Experiment on MFCC single-channel and MFCC dual-channel methods using audio recording data without noise and with noise. For audio recordings with noise using a noise variant from low to high. The test data also uses audio recordings with reduced noise with ANC. The feature analysis of each method uses k-mean clustering. The experimental results for the MFCC single-channel method are shown in Figure 7.

Experiments using the MFCC single-channel method to sample words without noise, the results are shown in Figure 7(a). The cluster center between the test and comparison samples has a high closeness value.

However, in the sample word experiment with noise, the cluster center of the test sample and the comparison of the proximity values are low, as shown in Figure 7(b).

The third test of variance used ANC to reduce the noise on the word sample before extraction. The result is that the centroid values in the test word sample are closer, as shown in Figure 7(c). The experiment for the MFCC dual-channel method used a word sample without noise, a word sample with noise, and a word sample with noise reduced by ANC. The test results are shown in Figure 8.

The experiment for the MFCC dual-channel method uses a word sample without noise value of the cluster centroid on channels 1 and 2 has a high closeness to the comparison word sample as shown in Figure 8. In the experiment of the MFCC dual-channel method for the word sample without noise, in channels 1 and 2, the proximity of the cluster center to the comparison center is high, as shown in Figure 8(a). As for the word sample with noise, the results on channel 1 are closer to the comparison than channel 2, as shown in Figure 8(b). In the test of the word sample with reduced noise using ANC, the result is that the cluster center between the test word sample and the comparison word sample is high, as shown in Figure 8(c).

The experiment used data with noise variance from 0 dB to -16 dB. The models tested are MFCC single-channel and MFCC dual-channel, the accuracy values, as shown in Figure 9. SNR variance in word samples to determine the performance of MFCC single-channel and MFCC dual-channel methods. Word samples with SNR -10 dB and -16 dB without being reduced by ANC, the results for the MFCC single-channel method have accuracy values of 57.5% and 47.5%. Meanwhile, with the MFCC dual-channel method, the accuracy values are 82% and 76.2%. Based on these experiments, showing the MFCC dual-channel method is more resistant to noise than the MFCC single-channel method, the experimental results, as shown in Figure 9(a).

The high noise in the word sample affects the accuracy the higher the noise value, the lower the accuracy value. To increase the accuracy value by separating the bandwidth on mel scale (MFCC dual-channel) and reducing noise (ANC) in the sample words before extraction. The experiments with SNR -16 dB on word samples reduce by ANC, the results with the MFCC single-channel method with an accuracy of 82.5%, while the MFCC dual-channel method with an accuracy value of 83.75%. The amount of noise reduction with ANC shows a proportional increase in the value of accuracy. Tests using high-quality recordings (without noise), the results for the MFCC single-channel method have an accuracy value of 92.5%, the MFCC dual-channel method has an accuracy value of 97.5%, as shown in Figure 9(b).

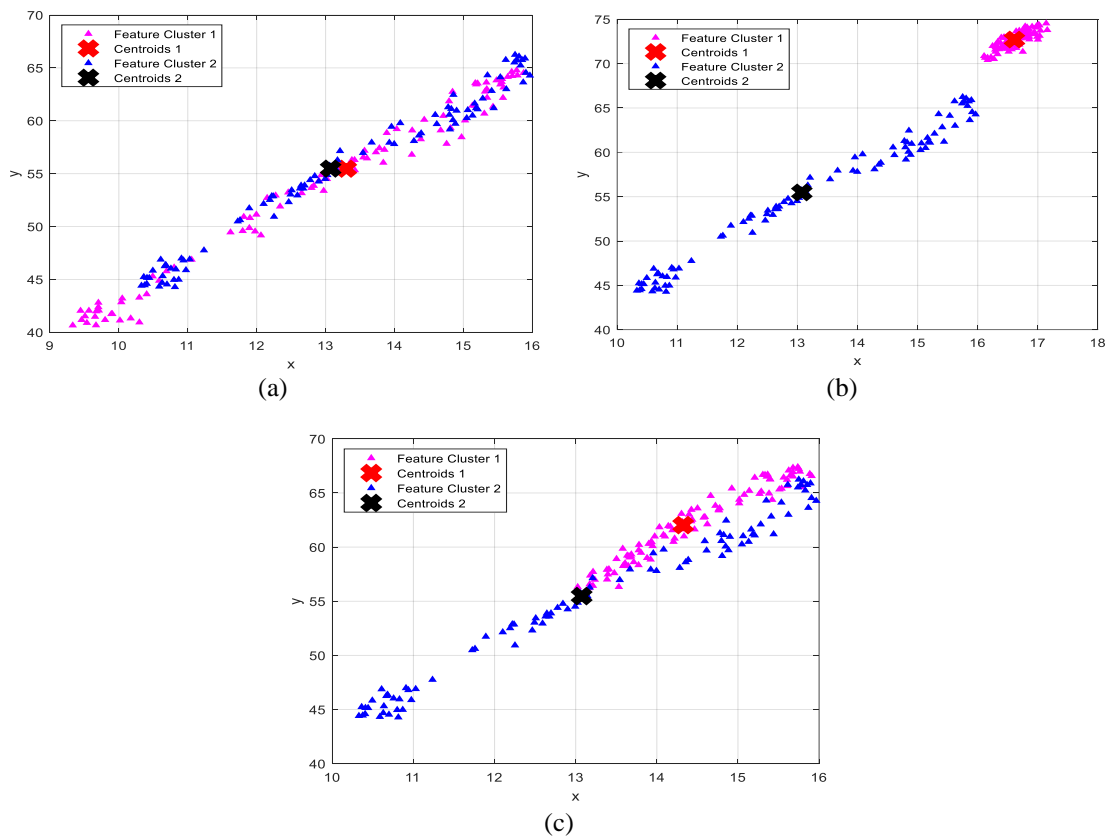

Figure 7. MFCC single-channel; (a) extracted word sample without noise, (b) the extraction result for the noise word sample is -6 dB, (c) extraction result for -6 dB noise word sample reduced by ANC

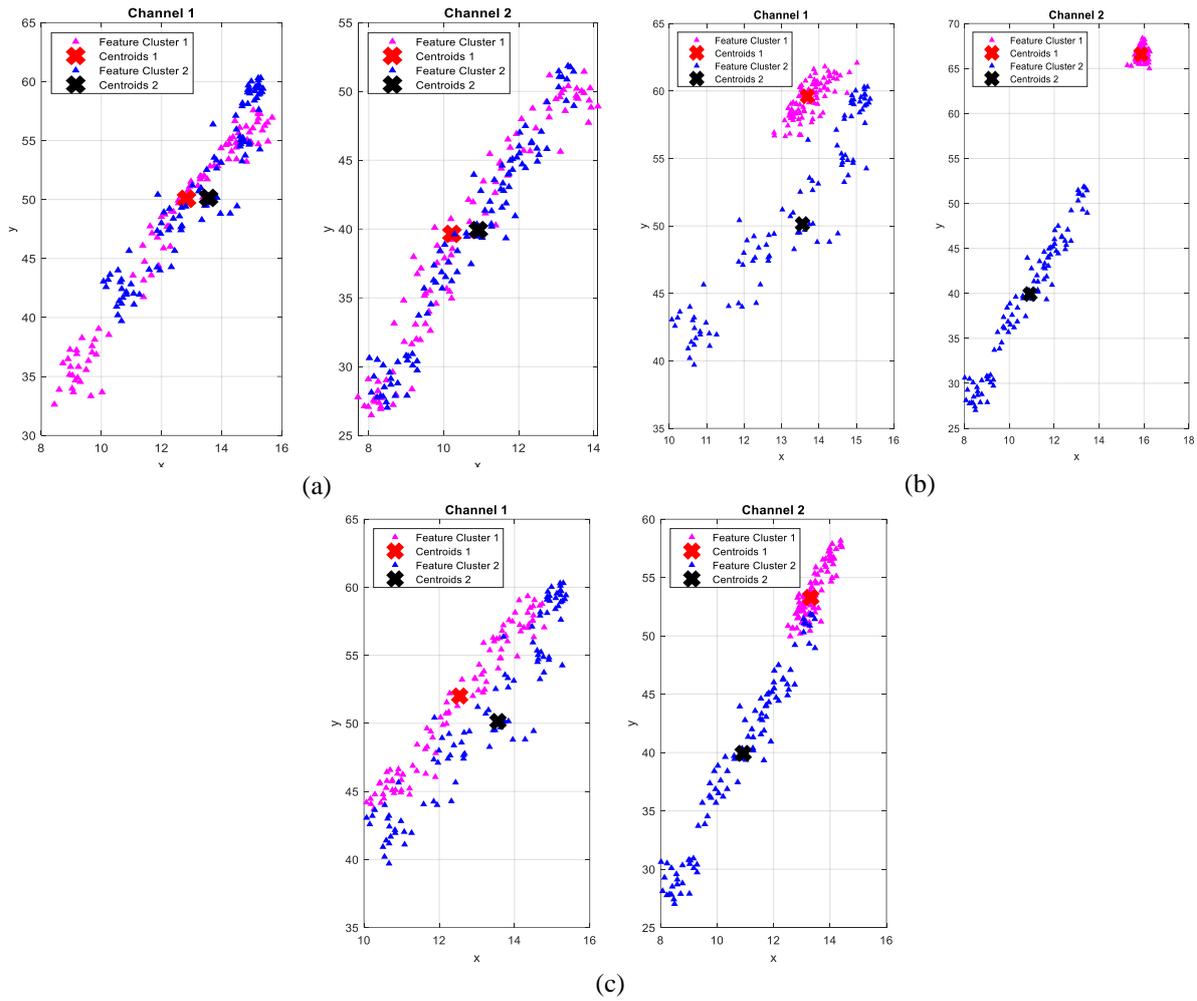

Figure 8. MFCC dual-channel method; (a) extracted word sample without noise; (b) the extraction result for the noise word sample is -6 dB; (c) extraction result for -6 dB noise word sample reduced by ANC

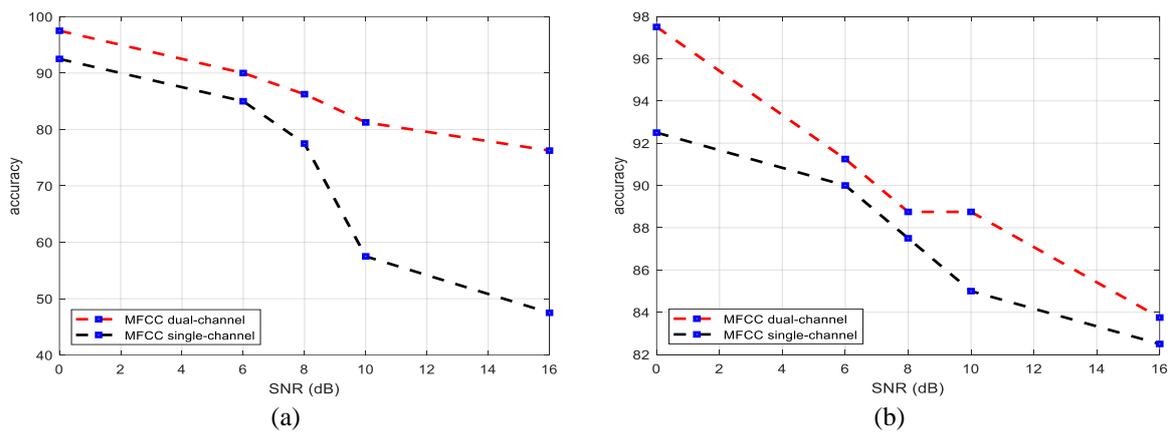

Figure 9. MFCC single-channel and MFCC dual-channel methods; (a) without ANC and (b) with ANC

## 4. CONCLUSION

Based on the experiment, it is concluded that the recorded noise affects the accuracy. High noise causes low accuracy. In the sample words with SNR -16 dB using the MFCC single-channel method with 47.5% accuracy and the MFCC double-channel method with 76.25% accuracy. The use of ANC can increase the accuracy of the MFCC single-channel by 82.5% and MFCC dual-channel method by 83.75%. In audio recording without noise for the MFCC single-channel method, the accuracy value is 92.5%. The MFCC dual-channel accuracy is 97.5%. Based on the test results, the MFCC dual-channel method has higher accuracy than the MFCC single-channel for recording with or without noise. The use of the mel scale separation method in MFCC dual-channel can increase the accuracy value.

## BIOGRAPHIES OF AUTHORS

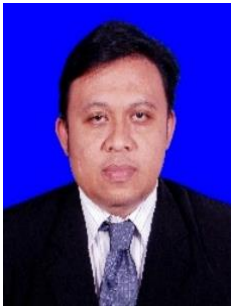

**Roy Rudolf Huizen** received his Bachelor of Engineering in Electrical Engineering (1999) from Universitas Semarang (USM) Semarang, Central Java. Master of Electrical Engineering (2006) and Doctor of Computer Science (2018) from Universitas Gadjah Mada (UGM) Yogyakarta, Indonesia. Lecturer and researcher at the Department of Electrical Engineering, Universitas Semarang (USM) from 2000 to 2008, and at the Faculty of Informatics and Computers, Institut Teknologi dan Bisnis STKOM Bali, from 2008 to the present. Research interest in signal processing and digital forensics.

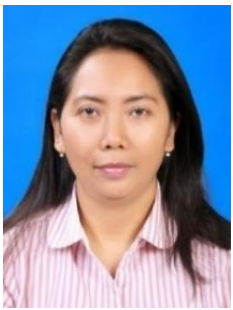

**Florentina Tatrin Kurniati** received the Bachelor of Engineering degree in Civil Engineering, from Universitas Semarang (USM), Central Java, Indonesia (2000) and Master of Engineering in Informatics from Universitas Atma Jaya Yogyakarta (UAJY), Yogyakarta, Indonesia (2015). Since 2008 she has been a lecturer and researcher at the Faculty of informatics and computer, Institut Teknologi dan Bisnis STKOM Bali, Indonesia. She is interested in adaptive noise cancellation, pattern recognition and digital forensics.